\documentclass[aps,twocolumn,showpacs,nofootinbib,superscriptaddress]{revtex4}
\usepackage{graphicx}
\usepackage{amsmath}
\usepackage{amsfonts}
\usepackage{amssymb}
\usepackage{hyperref}

\newtheorem{theorem}{Theorem}
\newtheorem{lemma}{Lemma}

\newcommand{\Zs}{\langle Z_1, \ldots, Z_n \rangle}
\newcommand{\Xs}{\langle X_1, \ldots, X_n \rangle}
\newcommand{\proof}{{\bf Proof:}}
\newcommand{\QED}{\hfill QED}

\newcommand{\ket}[1]{|{#1}\rangle}
\newcommand{\bra}[1]{\langle{#1}|}

\begin{document}

\title{Classicality in discrete Wigner functions}

\author{Cecilia Cormick}
\affiliation{Departamento de F\'\i sica ``Juan Jos\'e Giambiagi'', FCEyN UBA,
Pabell\'on 1, Ciudad Universitaria, 1428 Buenos Aires, Argentina}

\author{Ernesto F. Galv\~{a}o}
\affiliation{Perimeter Institute for Theoretical Physics \\31
Caroline Street North, Waterloo, Ontario, N2L 2Y5, Canada}

\author{Daniel Gottesman}
\affiliation{Perimeter Institute for Theoretical Physics \\31
Caroline Street North, Waterloo, Ontario, N2L 2Y5, Canada}

\author{Juan Pablo Paz}
\affiliation{Departamento de F\'\i sica ``Juan Jos\'e Giambiagi'', FCEyN UBA,
Pabell\'on 1, Ciudad Universitaria, 1428 Buenos Aires, Argentina}
\affiliation{Theoretical Division, Los Alamos National Laboratory, MS B288, Los Alamos,
NM 87545, USA}

\author{Arthur O. Pittenger}
\affiliation{Department of Mathematics and Statistics \\University
of Maryland, Baltimore County \\Baltimore, MD 21250, U.S.A.}

\date{\today}
\begin{abstract}
Gibbons \textit{et al.} [Phys. Rev. A \textbf{70}, 062101(2004)]
have recently defined a class of discrete Wigner functions $W$ to
represent quantum states in a Hilbert space with finite dimension.
We show that the only pure states having non-negative $W$ for all
such functions are stabilizer states, as conjectured by one of us
[Phys. Rev. A \textbf{71}, 042302 (2005)]. We also show that the
unitaries preserving non-negativity of $W$ for all definitions of
$W$ form a subgroup of the Clifford group. This means pure states
with non-negative $W$ and their associated unitary dynamics are
classical in the sense of admitting an efficient classical
simulation scheme using the stabilizer formalism.

\end{abstract}
\date{\today}
\maketitle

\section{Introduction}

Continuous-variable Wigner functions $W(q,p)$ have been used for a long time
to represent quantum systems in phase space \cite{Wigner32, HilleryOSW84}.
The Wigner function $W(q,p)$ is an alternative complete description of quantum
states which behaves almost like a phase-space probability density. Not only is
it real-valued and normalized but it also yields the correct value of the
probability density for the quadrature $a\hat Q+b\hat P$ when integrated along
the phase-space line $aq+bp$. However, unlike probability densities, the Wigner
function can assume negative values for some quantum states. This negativity of
the Wigner function has been considered a defining signature of non--classicality
(or quantum coherence and interference)
\cite{Zurek03, PazZ00}.

In quantum information science we usually deal with systems with a
space of states with a finite dimension $d$. For example, for a
system of $n$ qubits the dimension of the (Hilbert) space of
states is $d=2^n$. For such systems, various discrete analogues of
the Wigner function have been proposed \cite{Buot74, HannayB80,
CohenS86, Feynman87, Wootters87, GalettiP88, GibbonsHW04} and used
to investigate a variety of interesting problems connected with
quantum computation such as the phase-space representation of
quantum algorithms \cite{MiquelPS02}, separability
\cite{PittengerR05}, quantum state tomography
\cite{Wootters04,PazRS04}, teleportation \cite{KoniorczykBJ01,
Paz02}, decoherence in quantum walks \cite{LopezP03},
 and error correction \cite{PazRS04b}.
Here we shall concentrate on a class of discrete Wigner functions $W$ introduced
recently by Gibbons \textit{et al.} \cite{GibbonsHW04}. This elegant approach seems
to be a potentially powerful tool to establish connections between phase-space
techniques and problems in quantum information and foundations of quantum mechanics.

In this paper we study the set of states with non-negative discrete Wigner functions
$W$ for all functions in the class proposed by \cite{GibbonsHW04}, and the group of
unitaries that preserve non-negativity of $W$. Our first result is a complete
characterization of the set of quantum states having non-negative discrete Wigner
functions $W$. This is done by proving a conjecture presented by one of us in
\cite{Galvao05} (a related discussion in a somewhat different context,
using concepts in high-dimensional geometry appeared in
\cite{Bengtsson04,BengtssonE04}). Our proof is elementary and constructive, and
shows that the only pure states with non-negative $W$ are stabilizer states, i.e. simultaneous eigenstates of generalized Pauli operators \cite{Gottesman97}. We
then study the group of unitaries which preserve non-negativity of $W$, and prove
that they form a subgroup of the Clifford group. This means such states and unitaries
are classical in the sense of allowing for an efficient classical simulation
scheme using the stabilizer formalism.

The paper is organized as follows. In Section II we review the discrete Wigner
functions $W$ of \cite{GibbonsHW04}. In Section III we characterize the states
with non-negative $W$, in Section IV we discuss positivity-preserving unitary
dynamics in phase space and in Section V we summarize our results.

\section{Discrete Wigner functions} \label{sec:wignerdef}

In this section we review the class of discrete Wigner functions proposed in
\cite{GibbonsHW04} and discuss some of their features.

Let us assume that we are describing a quantum state whose Hilbert
space dimensionality $d$ is a power of a prime number $p$
($d=p^n$). In such cases one can introduce a phase-space grid with
$d\times d$ points and label the position and momentum coordinates
$(q,p)$ with elements of the finite Galois field $GF(p^n)$
\cite{LidlN86}. At first the use of elements of $GF(p^n)$ for both
phase-space coordinates could be seen as an unnecessary
complication, but it turns out to be an essential step. The reason
is that by doing this we can endow the phase-space grid with the
same geometric properties as the ordinary plane. For example, in
the finite $d\times d$ grid we can define lines as solutions to
linear equations of the form $aq+bp=c$ [where all elements and
operations in this equation are in $GF(p^n)$]. Each line will then
consist of exactly $d$ points of the grid. The field structure of
$GF(p^n)$ ensures the validity of properties such as: (i) there is
only one line joining any given two points, (ii) two lines are
either parallel (i.e. with no points in common) or they intersect
at a single point. Moreover, it is possible to show that a set of
$d$ parallel lines (which we will call a
\textit{striation}\cite{GibbonsHW04}) is obtained by varying the
parameter $c$ in the equation $aq+bp=c$. Finally, the number of
different striations turns out to be $(d+1)$. The complete set of
$(d+1)$ striations has been studied for a long time in discrete
geometry, where it is called a finite affine plane
\cite{Wootters04b, LidlN86}. We will label the striations with an
index $\kappa=1,\ldots ,d+1$ and the lines within a striation with
an index $j=1,\ldots,d$. In this way the $j$-th line belonging to
the $\kappa$--th striation will be denoted as
$\lambda^{(\kappa)}_j$.

A discrete phase space with the above properties was used by
Gibbons, Hoffman and Wootters in \cite{GibbonsHW04} to define a
class of discrete Wigner functions. As mentioned above, the
crucial property of the continuous Wigner function is that its
integral along any line $\lambda$ is equal to the expectation
value of a projection operator $\hat{P}_\lambda$, i.e. a
probability. This essential feature is generalized to the discrete
case in a straightforward way: every line in the $d\times d$
phase-space grid is associated to a rank one projection operator.
As noted in \cite{GibbonsHW04}, this association cannot be
arbitrary and must obey some simple geometric constraints. For
example, we can define a set of $d\times d$ unitary operators
$\hat{T}(q,p)$ acting on the Hilbert space that faithfully
represent discrete phase-space translations. For the association
between lines and states to respect covariance under translations
we must impose that the quantum state associated to a translated
line should be identical to the state obtained by acting with the
operator $\hat{T}(q,p)$ on the original state. This covariance
constraint can be used to show the validity of some very
significant properties: a) the states associated to parallel lines
must be orthogonal; b) the overlap between states associated to
non-parallel lines must be equal to $1/d$. This is important and
implies that the $(d+1)$ phase-space striations must be associated
to an equal number of mutually unbiased bases (MUB), i.e. bases
\begin{equation}
MUB^{(\kappa)}=\{|\phi^{(\kappa)}_1\rangle,\ldots, |\phi^{(\kappa)}_d\rangle\}
\label{MUB}
\end{equation}
such that
\begin{equation}
|\langle
\phi^{(\kappa')}_{j'}|\phi^{(\kappa)}_j\rangle|^2=\frac{1}{d}
(1-\delta_{\kappa,\kappa'})+\delta_{\kappa,\kappa'} \delta_{j,j'}.
\label{eqmubcond}
\end{equation}

As we see, mutually unbiased bases are orthonormal bases picked in such a way
that any state in one basis is an equal--amplitude superposition of all the
states of any other basis. A complete set of $(d+1)$ MUB is known to exist
if the dimensionality of the space of states is a power of a prime number.
In such case, many constructions of MUB have been proposed \cite{Ivanovic81,
WoottersF89, LawrenceBZ02, BandyopadhyayBRV02, PittengerR04}. It has been shown
that a complete set of $(d+1)$ MUB for $d$-dimensional systems can be chosen
to consist solely of stabilizer states, i.e. simultaneous eigenstates of sets
of (generalized) Pauli operators \cite{LawrenceBZ02,BandyopadhyayBRV02,PittengerR04}.

The defining feature of the discrete Wigner functions of
\cite{GibbonsHW04} is the association between MUB and striations
in the discrete phase space. As discussed in \cite{GibbonsHW04}
this can be done in a variety of ways and each defines a different
\textit{quantum net} \cite{GibbonsHW04}, which will result in a
different definition of the discrete Wigner function $W$. In this
paper we propose a notion of classicality of quantum states which
is based on non-negativity of $W$ for \textit{all} quantum nets
obtainable from a fixed complete set of MUB. It should be noted,
however, that there have been proposals of criteria to narrow down
the choice of quantum nets: in \cite{PazRS04b} the criterion is
covariance under the so-called discrete squeezing operator; and in
\cite{PittengerR05} the net is chosen so as to enforce a natural
relation between a separable state's $W$ and the $W$ of its
subsystems.

The quantum net is defined by associating each line
$\lambda^{(\kappa)}_j$ in striation $\kappa$ to a projector
$\hat{P}_j^{(\kappa)}=|\phi^{(\kappa)}_j\rangle\langle
\phi^{(\kappa)}_{j} |$ onto a basis state  of basis $\kappa$.
Having fixed a quantum net, the discrete Wigner function is
uniquely defined by imposing the condition that the sum of its
values along any line should be equal to the expectation value of
the projector corresponding to that line (see \cite{GibbonsHW04}
for details). The resulting Wigner function at any phase-space
point $\alpha=(q,p)$ can then be shown to be
\begin{eqnarray}
W_{\alpha}&=&{\rm Tr}\left(\hat{\rho} \hat{A}(\alpha)\right),\\
\hat{A}(\alpha)&=&\frac{1}{d}\left(
\sum_{\lambda^{(\kappa)}_j \ni \alpha} \hat{P}^{(\kappa)}_j - \mbox{$1
\hspace{-1.0mm}  {\bf l}$}  \right),\label{eqpointop}
\label{eqwgen}
\end{eqnarray}
where the sum is over projectors associated with all lines $\lambda^{(\kappa)}_j$ containing point $\alpha$. The construction of the striations guarantees that the sum above will contain exactly one projector from each basis.

The operators $\hat{A}(\alpha)$ are known as \textit{phase-space point operators} and form a
complete basis for the space of operators, which is orthogonal in the Schmidt inner
product (i.e. ${\rm Tr}\left(\hat{A}(\alpha)\hat{A}(\beta)\right)=\delta_{\alpha,\beta}/d$).
We can rewrite the expression for the Wigner function at phase-space point $\alpha$ using
the probabilities associated with the projectors $\hat{P}^{(\kappa)}_j$:
\begin{equation}
p^{(\kappa)}_j\equiv{\rm Tr}\left(\hat{\rho} \hat{P}^{(\kappa)}_j\right).
\label{defprob}
\end{equation}
In terms of these probabilities, the Wigner function at the point $\alpha$ takes the form
\begin{equation}
W_{\alpha}=\frac{1}{d} \left( \sum_{\lambda^{(\kappa)}_j \ni \alpha}
p^{(\kappa)}_j - 1  \right).
\label{eqwprob}
\end{equation}

The discrete Wigner function $W$ can be shown to have many of the
features of the continuous Wigner function $W(q,p)$
\cite{GibbonsHW04}: it is real (but can be negative), normalized,
and its values are obtained through eq. (\ref{eqwprob}) from
measurements onto MUB. Here the MUB projectors play the role that
the quadratures $a\hat{Q}+b\hat{P}$ play in $W(q,p)$, forming a
particularly symmetric set of observables whose measurement
results completely characterize the state (in a process known as
\textit{quantum tomography}). For a discussion of further
properties of $W$ see \cite{GibbonsHW04, PazRS04b, PittengerR04}.

In the discussion that follows we will often be representing
quantum states using the probabilities $p^{(\kappa)}_j$. As the
projectors $\hat{P}^{(\kappa)}_j$ form an over--complete basis for
the space of density matrices, these probabilities completely
characterize the state. Since for any striation $\sum_j
p^{(\kappa)}_j=1$, there are only $(d-1)$ independent
probabilities for each basis, resulting in a total of
$(d-1)\cdot(d+1)=(d^2-1)$ independent probabilities, exactly the
number of real parameters necessary to describe a general
normalized mixed quantum state in $d$-dimensional Hilbert space.
Each quantum state is represented by a point $\vec{p}$ in this
$(d^2-1)$-dimensional probability space.

As mentioned above, for power-of-prime $d$ it is possible to build
a complete set of $(d+1)$ MUB using only stabilizer states, i.e.
joint eigenstates of generalized Pauli operators. Let us discuss
more explicitly such constructions for the case $d=2^n$, i.e. $n$
qubits (see \cite{LawrenceBZ02} for more details). In order to
define a complete set of $(2^n+1)$ MUB we start by partitioning
the $(4^n -1)$ Pauli operators (excluding the identity) into
$(2^n+1)$ sets $S_i$ of $(2^n-1)$ Pauli operators each. We will
require that the Pauli operators in each set $S_i$ be mutually
commuting, but otherwise the partitioning can be completely
arbitrary. If we add the identity and a $\pm1$ phase to the Pauli
operators in each set $S_i$, each will form a maximal Abelian
subgroup of the Pauli group. The joint eigenstates of each such
set $S_i$ form a basis for the Hilbert space, and due to
properties of the Pauli operators the $(2^n+1)$ bases thus defined
can be shown to be mutually unbiased \cite{LawrenceBZ02}.

The phase-space construction provides a natural procedure for
partitioning the Pauli group into disjoint, mutually commuting
sets. The idea, which is worth reviewing here, was described in
\cite{GibbonsHW04} and further elaborated in \cite{PazRS04b}.
Pauli operators represent phase-space translations and can be
labelled using binary $n$--tuples $\vec{p}$ and $\vec{q}$
($n$--tuples $\vec{q}$ and $\vec{p}$ contain the coordinates of
the field elements $q$ and $p$ in a given basis as described
below). Each Pauli operator can be written as
\begin{equation}
\hat{T}(\vec q,\vec p) = \prod_{i=0}^{n-1}\hat{X}_i^{q_i}\hat{Z}_i^{p_i}e^{i\frac{\pi}{2}q_i \cdot p_i},
\end{equation}
where $\hat{X}_i$ and $\hat{Z}_i$ stand for the Pauli operators on qubit $i$, and the phase is chosen so as to make the operators Hermitian. The definition above will be written in shorthand as
\begin{equation}
\hat{T}(\vec q,\vec p) =  \hat{X}^{\vec q} \hat{Z}^{\vec p} e^{i
\frac{\pi}{2} \vec {q} \cdot \vec {p}}.
\end{equation}
The condition for two Pauli operators to commute turns out to be
\begin{equation}
[\hat{T}(\vec q,\vec p),\hat{T}(\vec {q^\prime},\vec {p^\prime})]=0 \text{ iff }
\vec{q}\cdot\vec{p^\prime}-\vec{p}\cdot\vec{q^\prime}=0 \quad ({\rm mod} ~2). \label{eqcomrel}
\end{equation}

Let us consider a set of $(d-1)$ Pauli operators
\begin{equation}
S_{(\vec a,\vec b)}=\left\{ \hat{T}(\vec {a} M^j, \vec {b} \widetilde {M}^j)
\quad j = 0, 1, \ldots, d-2\right\} ,
\label{eqsetPaulis}
\end{equation}
where $M$ is an arbitrary binary matrix, $\widetilde {M}$ is its
transpose and $\vec a$, $\vec b$ are binary $n$--tuples. Any two
operators of this set commute. It is interesting to note that
$S_{(\vec a,\vec b)}$ forms a maximal Abelian subgroup of the
Pauli group if and only if $M$ is a generating element of the
matrix representation of the field $GF(2^n)$. This can be seen as
follows: the product of two elements of $S_{(\vec a,\vec b)}$ is
itself an element of this set (up to a sign) iff the matrix $M$ is
such that for every pair of integers $j,j'$, there is a third
integer $j''$ such that $M^j+M^{j'}=M^{j''}$ [where $0 \le j, j',
j'' \le (d-2)$ and $j \neq j'$]. Moreover, for the set to have
exactly $(d-1)$ different elements, the matrix $M$ should be such
that all powers $M^j$ for $j=0,\ldots,(d-2)$ are nonzero and
different from each other. For $M$ satisfying these conditions, it
can be seen that $M^{d-1}=\mbox{$1 \hspace{-1.0mm}  {\bf l}$}$.
Therefore, the elements of the set $\{{\bf 0}, \mbox{$1
\hspace{-1.0mm} {\bf l}$}, M, M^2,\ldots, M^{d-2}\}$ form a finite
field, and we see that the matrix $M$ and its powers form a matrix
representation of $GF(2^n)$. A possible choice for $M$, used in
\cite{PazRS04b}, is the so--called ``companion matrix'' of the
primitive polynomial which defines the product rule in the field.
With such a matrix we can build $(d+1)$ disjoint sets of commuting
Paulis of the form $S_{(\vec a,\vec b)}$ by choosing the binary
$n$--tuples $(\vec a,\vec{b})$ as explained below.

The association between each phase-space point $(q,p)$ and a Pauli
operator $\hat{T}(\vec{q},\vec{p})$ must respect the covariance of
the construction under phase space translations. This is done as
follows: the line formed by all phase-space points satisfying the
equation $bq+ap=c$ is invariant under phase-space translations of
the form $q'=q+a\omega^j$, $p'=p+b\omega^j$ (where $\omega$ is a
generating element of the field). To this phase-space translation
we must associate an operator acting in Hilbert space. The natural
identification is to associate this with the operator
$\hat{T}(\vec {a} M^j, \vec {b} \widetilde {M}^j)$. Here, the
choice of $n$--tuples $\vec{a}$ and $\vec{b}$ is arbitrary. The
important point is that once this choice is made [i.e., once we
arbitrarily assign two $n$--tuples to the point $(a,b)$] we
repeatedly apply the matrix $M$ ($\widetilde{M}$) to the position
(momentum) coordinates to obtain the $n$--tuples parametrizing the
Pauli operators associated to the other phase-space points
\cite{PazRS04b}. In summary, this construction associates an
operator $\hat{T}(\vec{q},\vec{p})$ to every phase space point
$(q,p)$ in such a way that the elements of the Abelian subgroup
$S_{(\vec a,\vec b)}$ are associated to points in phase space that
belong to the \textit{ray} defined by the equation $bq+ap=0$ (a
\textit{ray} is defined as a line that contains the origin). In
\cite{PazRS04b} it was shown that by varying the $n$--tuples
$(\vec{a},\vec{b})$ one can construct only $(d+1)$ different sets
$S_{(\vec{a},\vec{b})}$. If we define the two $n$--tuples
$\vec{1}\equiv(1000\cdots0)$ and $\vec{0}\equiv(0000\cdots0)$,
these maximal mutually commuting sets of Pauli operators can be
conveniently built by choosing
$(\vec{a},\vec{b})=(\vec{1},\vec{0)}$ (which we will associate
with the horizontal striation),
$(\vec{a},\vec{b})=(\vec{0},\vec{1)}$ (the vertical striation) and
$(\vec{a},\vec{b})=(\vec{1},\vec{b)}$ for $\vec{b}\neq \vec{0}$
(the other striations).

Thus, the mapping between striations and MUB is naturally
determined by the phase-space construction, as lines which are
invariant under the transformations $q'=q+a\omega^j$,
$p'=p+b\omega^j$ must be associated to states which are invariant
under the corresponding transformations in Hilbert space, that is,
the translation operators in $S_{(\vec a,\vec b)}$. Therefore, the
lines of the form $bq+ap=c$ must be associated to common
eigenstates of the set $S_{(\vec a,\vec b)}$. However, there is no
criterion telling us how to associate each line in a striation
with a projector in the corresponding basis. We can count the
number of possible quantum nets as follows: for the ray of a given
striation there are $d$ possible projectors to choose from; once
this choice has been made the condition of covariance under
translations determines which projector should be associated to
each of the other lines in the same striation. As there are
$(d+1)$ rays, the number of possible associations between lines
and projectors is $d^{d+1}$, each of which defines a different
quantum net, leading to a different definition of the Wigner
function. The projectors associated to each of the lines in the
vertical and horizontal striations can be chosen in such a way
that the coordinates of each line correspond to the eigenvalues of
the Paulis generating the set (the single qubit Paulis $\hat{Z}$
and $\hat{X}$, respectively). Then, there are still $d^{d-1}$
possible quantum nets, each of them given by a particular choice
of projectors to be associated to the rays of the remaining
\textit{oblique} striations.

There is a closely related methodology for constructing the Wigner
functions for general dimension $d = p^n$ which emphasizes the
link between the exponents of the $\hat{Z}$ and $\hat{X}$
operators and the finite geometry of $V_2[GF(p^n)]$, the
two-dimensional vector space over the field $GF(p^n)$. One defines
lines, rays and striations in this two dimensional space and then,
using the properties of the algebraic field extension, defines an
isomorphism with $V_{2n}[GF(p)]$. Vectors in this second space
serve as exponents of the $\hat{Z}$ and $\hat{X}$ operators, and
the commuting classes of generalized Pauli matrices correspond
precisely to parallel lines in a striation in the first vector
space. Details of this approach and a methodology for assigning
projections to lines are given in \cite{PittengerR04}.

\section{States with non-negative Wigner functions $W$}

Following \cite{Galvao05}, let us now characterize the set of
states having non-negative discrete Wigner functions $W$
simultaneously in all definitions proposed by Gibbons \textit{et
al.} \cite{GibbonsHW04} for power-of-prime dimension $d$.

\begin{description}
\item[Definition:] The set $C_d$ is defined as the set of (pure or mixed) density
matrices of systems in a $d$-dimensional Hilbert space having non-negative discrete
Wigner function $W$ in all phase-space points \textit{and} for all definitions of $W$
using a fixed set of mutually unbiased bases.
\end{description}
By definition, the set $C_d$ is specified as the intersection of a
number of half-spaces in the $(d^2-1)$-dimensional $\vec{p}$-space. From (\ref{eqwprob}) it can be seen that each half-space inequality is of the form
\begin{equation}
 \sum_{\lambda^{(\kappa)}_j \ni \alpha} p^{(\kappa)}_j \ge 1,
\end{equation}
where the probabilities appearing in the sum are associated with the lines containing phase-space point $\alpha$, and hence depend on $\alpha$ and on the quantum net chosen. The intersection of the half-spaces defined by these inequalities is a convex polytope in $\vec{p}$-space, given in an H-description (H
standing for ``Half-space''). Any convex polytope also admits an alternative V-description
(V for ``Vertices''), consisting of the list of vertices whose convex hull defines the polytope.

Galv{\~a}o showed that for $d \le 5$, the H-polytope $C_d$ has a V-description whose
vertices are the MUB projectors \cite{Galvao05}, and conjectured this would also be
true for general power-of-prime $d$. A geometrical argument showing the validity of
this conjecture was given in \cite{Bengtsson04, BengtssonE04}. Let us now provide a constructive, analytical proof.
\begin{theorem}
For any power-of-prime Hilbert space dimension $d$, the H-polytope $C_d$ is equivalent to the V-polytope $C_v$ having the MUB projectors as vertices. \label{thm:statics}
\end{theorem}
\proof

Let us prove the theorem by first showing that the V-polytope $C_v$ is contained in $C_d$, and then the converse. From \cite{GibbonsHW04} we know that the Wigner function for any MUB projector
$\hat{P}_{j}^{(\kappa)}=|\phi_{j}^{(\kappa)}\rangle
\langle\phi_{j}^{(\kappa)}|$ is non-negative. Since
the Wigner function depends linearly on the density
matrices, $W$ is non-negative also for any state in the convex
hull of the $\hat{P}_{j}^{(\kappa)}$. This shows that any state in
$C_v$ is also in $C_d$, as we wanted to prove.

Let us now prove the converse, i.e. that polytope $C_v$ is contained in polytope $C_d$. What is required now is to show that any state $\hat{\rho} \in C_d$ can be written as a convex combination of the projectors $\hat{P}_{j}^{(\kappa)}$:
\begin{equation}
\hat{\rho}=\sum_{\kappa=1}^{d+1} \sum_{j=1}^d
c_{j}^{(\kappa)}\hat{P}_{j}^{(\kappa)},\label{eqrhoexp}
\end{equation}
with all $c_{j}^{(\kappa)}\ge0$. Note that the decomposition is
not unique.

Let us start by considering the general expression for Wigner
function $W$ at phase-space point $\alpha$ [eq. (\ref{eqwprob})].
Given a state $\hat{\rho}$ there is some Wigner function
definition which, at some point $\alpha$, evaluates to a minimum
value among all definitions and all points $\alpha$. This happens
when the expression for $W_{\alpha}$ is such that the sum
(\ref{eqwprob}) includes only the smallest probability
$p_{j}^{(\kappa)}$ from each MUB $\kappa$. Let us denote these
$W$-minimizing probabilities $p_{*}^{(\kappa)}\equiv
\min_j\{p_{j}^{(\kappa)}\}$. States in $C_d$ are those for which
all expressions of the form (\ref{eqwprob}) are non-negative, and
this happens if and only if the expression for $W$ involving only
the $p_{*}^{(\kappa)}$ is non-negative. In other words, a state
has non-negative $W$ in all definitions if and only if:
\begin{equation}
\sum_{\kappa=1}^{d+1}p_{*}^{(\kappa)}\ge 1.
\end{equation}
This is our hypothesis.

Any density matrix $\hat{\rho}$ can be expanded in terms of the projectors
$\hat{P}_{j}^{(\kappa)}=\ket{\phi_{j}^{(\kappa)}} \bra{\phi_{j}^{(\kappa)}}$ as in eq.
(\ref{eqrhoexp}), with real (but possibly negative) coefficients
$c_{j}^{(\kappa)}$. A first constraint on the coefficients
$c_{j}^{(\kappa)}$ comes from the requirement that
${\rm Tr}(\hat{\rho})=1$. Using property (\ref{eqmubcond}) of the MUB we
can compute the trace, obtaining
\begin{equation}
{\rm Tr}(\hat{\rho})=\sum_{\kappa=1}^{d+1}\sum_{j=1}^d
c_{j}^{(\kappa)}=1.
\label{eqtr1cond}
\end{equation}

Now let us use eq. (\ref{defprob}) to calculate $p_{j}^{(\kappa)}$
explicitly from eq. (\ref{eqrhoexp}), so as to obtain relations
between the coefficients $c_{j}^{(\kappa)}$ and the probabilities
$p_{j}^{(\kappa)}$:
\begin{eqnarray}
p_{j}^{(\kappa)}=\bra{\phi_{j}^{(\kappa)}} \hat{\rho} \ket{\phi_{j}^{(\kappa)}} = \sum_{\mu=1}^{d+1} \sum_{m=1}^d c_{m}^{(\mu)}\left| \langle\phi_{j}^{(\kappa)}|\phi_{m}^{(\mu)}\rangle\right|^2=\nonumber\\
=\sum_{\mu\neq \kappa} \sum_{m}c_{m}^{(\mu)}\left| \langle\phi_{j}^{(\kappa)}|\phi_{m}^{(\mu)}\rangle\right|^2 + \sum_{m}c_{m}^{(\kappa)} \left| \langle\phi_{j}^{(\kappa)}|\phi_{m}^{(\kappa)}\rangle\right|^2=\nonumber\\
=\sum_{\mu\neq \kappa}\sum_{m}c_{m}^{(\mu)}\frac{1}{d}+c_{j}^{(\kappa)},\quad \quad
\end{eqnarray}
where we have used the condition of mutual unbiasedness of the
bases [eq. (\ref{eqmubcond})]. Now we can use the trace condition
(\ref{eqtr1cond}) to rewrite this as
\begin{equation}
p_{j}^{(\kappa)}=c_{j}^{(\kappa)}+\frac{1}{d}-\frac{1}{d}\sum_{m}c_{m}^{(\kappa)}
\end{equation}
or
\begin{equation}
c_{j}^{(\kappa)}=p_{j}^{(\kappa)}-\frac{1}{d}+\frac{1}{d}\sum_{m}c_{m}^{(\kappa)}.
\end{equation}
Let us add $0=p_{*}^{(\kappa)}-p_{*}^{(\kappa)}$ to the right-hand side of the
equation above, to obtain
\begin{equation}
c_{j}^{(\kappa)}=\left(
p_{j}^{(\kappa)}-p_{*}^{(\kappa)}\right)+x^{(\kappa)}
\label{eqcfinal}
\end{equation}
with
\begin{equation}
x^{(\kappa)}\equiv
p_{*}^{(\kappa)}-\frac{1}{d}+\frac{1}{d}\sum_{m}c_{m}^{(\kappa)}.
\end{equation}

Eq. (\ref{eqcfinal}) tells us that each coefficient
$c_{j}^{(\kappa)}$ can be written as the sum of a non-negative
term $\left( p_{j}^{(\kappa)}-p_{*}^{(\kappa)}\right)$ plus a
(possibly negative) constant $x^{(\kappa)}$. We can show, however,
that the sum of those constants $x^{(\kappa)}$ has to be
non-negative. We do that by using the normalization condition
(\ref{eqtr1cond}) on eq. (\ref{eqcfinal}):
\begin{eqnarray}
\sum_{\kappa,j}c_{j}^{(\kappa)}=1 \Rightarrow \sum_{\kappa,j}p_{j}^{(\kappa)}-\sum_{\kappa,j}p_{*}^{(\kappa)}+\sum_{\kappa,j}x^{(\kappa)} =1\nonumber\\
\Rightarrow d+1-d\sum_{\kappa}p_{*}^{(\kappa)}+d\sum_{\kappa}x^{(\kappa)}=1\nonumber\\
\Rightarrow \sum_{\kappa}
x^{(\kappa)}=\sum_{\kappa}p_{*}^{(\kappa)}-1\label{eqxinp}.
\end{eqnarray}
Now remember that our hypothesis is that $\sum_{\kappa}
p_{*}^{(\kappa)}\ge 1$, which implies that $\sum_{\kappa}
x^{(\kappa)} \ge 0$. Let us now use this fact and expression
(\ref{eqcfinal}) to obtain an expansion of the density matrix
$\hat{\rho}$ in terms of the projection operators
$\hat{P}_{j}^{(\kappa)}$, but now with non-negative coefficients
only. Plugging eq. (\ref{eqcfinal}) into eq. (\ref{eqrhoexp}) we
obtain:
\begin{equation}
\hat{\rho}=\sum_{\kappa,j}
(p_{j}^{(\kappa)}-p_{*}^{(\kappa)}+x^{(\kappa)})\hat{P}_{j}^{(\kappa)}.
\end{equation}
Using the fact that $\sum_j \hat{P}_j^{(\kappa)}=\mbox{$1
\hspace{-1.0mm}  {\bf l}$}$, we can rewrite this as
\begin{equation}
\hat{\rho}=\sum_{\kappa,j}\left(p_{j}^{(\kappa)}-p_{*}^{(\kappa)}+\frac{x}{d+1}\right)\hat{P}_{j}^{(\kappa)}\label{eqnnconv}
\end{equation}
where we defined $x\equiv \sum_{\kappa} x^{(\kappa)}$. Note that
$p_{j}^{(\kappa)} \ge p_{*}^{(\kappa)}$ by definition, and our
hypothesis guarantees that $x\ge 0$. What we have now is then an
expansion of $\hat{\rho}$ in terms of the MUB projectors using
only non-negative coefficients. \QED
\medskip

Our proof above is constructive -- for any state in $C_d$ we can use equations (\ref{defprob}) and (\ref{eqxinp}) to obtain a convex
decomposition of the state in terms of the MUB projectors and their associated probabilities, given
by eq. (\ref{eqnnconv}). As noted in \cite{Galvao05}, some non-physical states
(i.e. described by non-positive Hermitian matrices) can have non-negative
Wigner functions in a single definition of $W$. Theorem \ref{thm:statics}
shows that imposing non-negativity of $W$ for all definitions of $W$ is
sufficient to guarantee that the set $C_d$ contains only physical states.

In the light of Theorem \ref{thm:statics} above, let us now discuss in which
senses states with non-negative Wigner functions are classical. We have defined
the set $C_d$ of states of a $d$-dimensional system with non-negative Wigner
function $W$ in all definitions. Theorem \ref{thm:statics} proves that the only
pure states in $C_d$ are the MUB projectors, which can always be chosen to be
stabilizer states, i.e. simultaneous eigenstates of Pauli operators
\cite{LawrenceBZ02, BandyopadhyayBRV02, PittengerR04}. The stabilizer formalism then provides us
with a way to represent pure states in $C_d$ using a number of bits which is
polynomial in the number of qubits \cite{NielsenC00}. This contrasts with general quantum states
whose classical description requires an exponential number of bits. We thus see
that states whose discrete phase-space description avoids Feynman's ``negative
probabilities'' \cite{Feynman87} are classical also in the sense of having a
classical-like short description.

\textit{Classicality witnesses} are observables whose expectation values, if negative, indicate some non-classical property such as entanglement (see \cite{Terhal02,KorbiczCWL05}). Our Theorem \ref{thm:statics} gives such an interpretation to the phase-space point operators $\hat{A}(\alpha)$: negative expectation values indicate (i.e. witness) non-classicality in the sense we discussed above.

The stabilizer formalism provides us with a framework in terms of which pure
states in the set $C_d$ have an efficient classical description. Other choices
of frameworks are possible, each choice resulting in a different set of quantum
states with efficient classical descriptions. One example are the (mixed)
separable density matrices of a collection of qubits, each of which has an
efficient classical description in terms of single-qubit pure states. An efficient
description, however, does not guarantee the existence of an efficient simulation
scheme for the dynamics; the dynamics of separable mixed states in NMR quantum
computation experiments provides us with an example of this problem
\cite{MenicucciC02}. In the next section we discuss the issue of simulability
of unitary dynamics within our set $C_d$.

Given these observations, it is not surprising that pure states in $C_d$ can
behave non-classically in other ways, that is, with respect to other frameworks.
For example, states in $C_d$ can be highly entangled, allowing for proofs of
quantum non-locality and contextuality.

\section{Unitaries preserving non-negativity of $W$} \label{secdynamics}

In continuous phase-space we can define classical unitaries as the
group of unitaries which preserve non-negativity of the Wigner
function $W(q,p)$. It has been shown that this group is formed by
all unitaries generated by Hamiltonians which are quadratic forms
in phase space \cite{HilleryOSW84}. In this section we obtain an
analogous result for our discrete Wigner functions $W$: using the
`classical' pure states in $C_d$, we define and characterize the
group of unitaries $\{U_c\}$ that map pure states in $C_d$ to
other pure states in $C_d$. In other words, we characterize the
group of `classical' unitaries $\{U_c\}$ that preserve
non-negativity of $W$ for all quantum nets obtained from a fixed
complete set of MUB.

The structure of the group $\{U_c\}$ may depend on the particular
complete set of MUB we choose to define our discrete Wigner
functions. We prove that for any MUB construction using Pauli
operators, the group $\{U_c\}$ is a subgroup of the Clifford group
(the group of unitaries mapping Pauli operators to Pauli operators
under conjugation \cite{NielsenC00}). For the particular
construction in \cite{GibbonsHW04, PazRS04b} we present some
unitaries in $\{U_c\}$ and discuss their action in phase space.

\subsection{$\{U_c\}$ is a subgroup of the Clifford group}

Let us consider the $(4^n-1)$ Pauli operators acting on the
$2^n$-dimensional Hilbert space of $n$ qubits (excluding the
identity). We can partition these $(4^n-1)$ Pauli operators into
$(2^n+1)$ sets $S_i$ of $(2^n-1)$ commuting Paulis each. The joint
eigenstates of the sets $S_i$ form a complete set of $(2^n+1)$
MUB, as discussed in section \ref{sec:wignerdef}. We can partition
the Paulis in many different ways; each such partitioning defines
a different complete set of MUB, which will be denoted as $B_j$.
In this section we show that the `classical' unitaries $\{U_c\}$
mapping MUB in a partition $B_i$ to MUB in the same partition form
a subgroup of the Clifford group.

The strategy is as follows. We will consider a slightly more
general problem, which is to characterize unitaries mapping MUB
defined by an arbitrary partition $B_1$ of Pauli operators to MUB
defined by a second partition $B_2$. A general unitary $U$ mapping
$B_1$ to $B_2$ will in particular map two bases in $B_1$ to two
other bases in $B_2$. Let us name them $(S_1 \in
B_1)\stackrel{U}{\mapsto} (S_2 \in B_2)$, $(T_1 \in B_1)
\stackrel{U}{\mapsto} (T_2 \in B_2)$. The first step is to prove
there are two Clifford unitaries $C_j (j=1,2)$ that map basis
$S_j$ to the computational ($Z$) basis, while mapping basis $T_j$
to the $X$-basis. These standard Clifford unitaries are the key to
the proof. This is because $U$ is Clifford if and only if
$\tilde{U} \equiv C_2UC_1^\dagger$ is Clifford. So it is enough to
show $\tilde{U}$ is Clifford (done in Theorem \ref{thm:auxxz}),
which is easier as by construction $\tilde{U}$ are unitaries that
preserve both the $Z$ basis and the $X$ basis.

With this more general result in hand, we can consider the case when the two partitions
are one and the same ($B_1=B_2$), and we will have what we wanted to prove, i.e. that
our `classical' unitaries $\{U_c\}$ are Clifford group operators.

We wish to show
\begin{theorem}
Let $U$ be a unitary transformation that maps a complete set of Pauli MUB $B_1$ to
a second complete set of Pauli MUB $B_2$.  Then $U$ is in the Clifford group, up to
a global phase. \label{thm:MUBequiv}
\end{theorem}

The first step involves proving the following Lemma:
\begin{lemma}
Let $S$ and $T$ be two maximal Abelian subgroups of the Pauli
group, with $S \cap T = \left\{ \mbox{$1 \hspace{-1.0mm}  {\bf
l}$}\right\}$. Then there exists a Clifford operation which maps
$S \mapsto \Zs$, $T \mapsto \Xs$.
\end{lemma}
\proof

Since $S$ and $T$ are maximal Abelian subgroups with trivial
intersection, it follows that no (non-identity) element of $T$
commutes with every element of $S$ (or vice-versa).
Let $\{M_i \quad i = 1, \ldots, n\}$ be a set of generators of
$S$. For any particular element $N \in T$, we can define the
syndrome $\vec{\sigma}(N)$ which is an $n$-tuple whose $i$-th component
is given by $\sigma_i = c(N, M_i) \quad i = 1, \ldots, n$. Here
$c(N, M) = 0$ if $N$ and $M$ commute and $c(N,M) = 1$ if $N$ and
$M$ anticommute.  Then it follows that if $N, N' \in T$, $N \neq
N'$, then $\vec{\sigma}(N) \neq \vec{\sigma}(N')$ (since otherwise
$\vec{\sigma}(NN') = \vec{\sigma}(N) + \vec{\sigma}(N') = \vec{0}$,
and $NN'$ would commute with every element of $S$).

In particular, since there are $2^n$ elements of $T$ and $2^n$
different possible values of $\vec{\sigma}$, it follows that each
value of $\vec{\sigma}$ is used exactly once.  Thus, we can choose
$N_i \in T$ such that $\vec{\sigma}(N_i) = \vec{e_i}$ (where
$\vec{e_i}$ is the vector that is $1$ in the $i$-th position and
$0$ elsewhere).  That is, $N_i$ anticommutes with $M_i$ and
commutes with $M_j$ ($i \neq j$).  The $N_i$'s are independent
(because their $\vec{\sigma}$ vectors are independent) and they
commute with each other (because $T$ is Abelian).  Therefore,
the set of $M_i$'s and $N_i$'s have the same
commutation/anticommutation relationships as the $Z_i$'s and the
$X_i$'s, so there exists a Clifford group operation that maps $M_i
\mapsto Z_i$ and $N_i \mapsto X_i$.  This provides the appropriate
map on $S$ and $T$.  \QED
\medskip

This Lemma can be adapted so it applies also to $d$-dimensional
registers, the main difference being that the syndrome function
$\vec{\sigma}(N)$ assumes values which are vectors modulo $d$ (see
\cite{Gottesman99b}).

An immediate consequence of this lemma is that for any complete
set of Pauli MUB $B$, we can choose any two of its bases,
represented by stabilizers $S$ and $T$, and find a Clifford group
operation that will map $B$ to another complete set of MUB
containing the bases $\Zs$ and $\Xs$; and in particular, this
Clifford group operation will map $S$ to $\Zs$ and $T$ to $\Xs$.

Therefore, if we have a general unitary $U$ that maps Pauli MUB
$B_1$ to Pauli MUB $B_2$, we can choose bases $S_1, T_1 \in B_1$
with $S_2 = U(S_1)$, $T_2 = U(T_1)$ (so $S_2, T_2 \in B_2$), and
then find Clifford operations $C_1$ and $C_2$ which map $C_1: S_1
\mapsto \Zs$, $C_1: T_1 \mapsto \Xs$, $C_2: S_2 \mapsto \Zs$,
$C_2: T_2 \mapsto \Xs$.  Then it follows that $C_2 U C_1^\dagger:
\Zs \mapsto \Zs$, $C_2 U C_1^\dagger: \Xs \mapsto \Xs$. Denoting $\tilde{U}=C_2 U C_1^\dagger$, it is easy to see that $\tilde{U}$ is a Clifford group operator iff $U$ is a Clifford group operation.  Thus, to prove
Theorem~\ref{thm:MUBequiv}, it will be sufficient to prove
\begin{theorem}\label{thm:auxxz}
If $\tilde{U}$ is a unitary operation which preserves both the $Z$ basis
and the $X$ basis (i.e., maps eigenstates of $\Zs$ to other eigenstates
of this set of operators, and the same is valid for eigenstates of $\Xs$), then
$\tilde{U}$ is a Clifford group operation, up to a global phase.
\end{theorem}
\proof

Since $\tilde{U}$ preserves the $Z$ basis, it has the form of a classical
gate with possibly some phases changed:
\begin{equation}
\tilde{U} \ket{\vec{z}}_Z = e^{i\phi(\vec{z})} \ket{\vec{g}(\vec{z})}_Z,
\end{equation}
with $\vec{g}(\vec{z})$ a permutation of the $2^n$ possible values of $\vec{z}$.

In terms of the $Z$ basis, we can expand elements of the $X$ basis
as follows:
\begin{equation}
\ket{\vec{x}}_X = \sum_{\vec{z}} e^{i\pi (\vec{x} \cdot \vec{z})} \ket{\vec{z}}_Z.
\end{equation}
Therefore,
\begin{eqnarray}
\tilde{U} \ket{\vec{x}}_X & = & \sum_{\vec{z}} e^{i(\pi \vec{x} \cdot \vec{z} + \phi(\vec{z}))}
\ket{\vec{g}(\vec{z})}_Z =\nonumber\\
& = & \sum_{\vec{z}} e^{i(\pi \vec{x} \cdot \vec{g}^{~-1}(\vec{z}) + \phi(\vec{g}^{~-1}(\vec{z})))}
\ket{\vec{z}}_Z. \label{eqn:UfromZ}
\end{eqnarray}
In order to preserve the $X$ basis, we need
\begin{equation}
\tilde{U} \ket{\vec{x}}_X = e^{i\theta(\vec{x})} \ket{\vec{h}(\vec{x})}_X = \sum_{\vec{z}} e^{i (\theta(\vec{x})
+ \pi \vec{h}(\vec{x}) \cdot \vec{z})} \ket{\vec{z}}_Z, \label{eqn:UfromX}
\end{equation}
where $\vec{h}(\vec{x})$ is a permutation of the values of
$\vec{x}$. Equating (\ref{eqn:UfromZ}) and (\ref{eqn:UfromX}), we
find
\begin{equation}
\pi \vec{x} \cdot \vec{g}^{~-1}(\vec{z}) +
\phi(\vec{g}^{~-1}(\vec{z})) = \theta(\vec{x}) + \pi
\vec{h}(\vec{x}) \cdot \vec{z}.\label{eq:phase1}
\end{equation}
This must be true for all $\vec{x}$ and $\vec{z}$.  Plugging in $\vec{x}=\vec{0}$,
we find
\begin{equation}
\phi(\vec{g}^{~-1}(\vec{z})) = \theta_0 + \pi \vec{h_0} \cdot \vec{z},
\label{eq:phig}
\end{equation}
where $\theta_0 = \theta(\vec{0})$ and $\vec{h_0} = \vec{h}(\vec{0})$.
Therefore
\begin{equation}
\pi \vec{x} \cdot \vec{g}^{~-1}(\vec{z}) = \pi
\left[\vec{h}(\vec{x})-\vec{h_0} \right] \cdot \vec{z} +
\left[\theta(\vec{x}) - \theta_0 \right].\label{eq:phase3}
\end{equation}
Of course, eqs. \ref{eq:phase1}-\ref{eq:phase3} are understood to
be modulo $2\pi$. It then follows that $\vec{g}^{~-1}(\vec{z})$
must be affine in $\vec{z}$:
\begin{equation}
\vec{g}^{~-1}(\vec{z}) = A\vec{z} + \vec{b},
\end{equation}
(where $A$ is an invertible $n \times n$ binary matrix) and thus,
by (\ref{eq:phig}), $\phi(\vec{z})$ is also affine in $\vec{z}$:
\begin{equation}
\phi(\vec{z}) = \pi \vec{c} \cdot \vec{z} + d,
\end{equation}
with $A \vec{c} = \vec{h_0}$ and $\pi \vec{b} \cdot \vec{c} + d =
\theta_0$.

Thus we find
\begin{equation}
\tilde{U} \ket{\vec{z}}_Z = e^{i(\pi \vec{c} \cdot \vec{z} + d)}
\ket{A^{-1}\vec{z} - A^{-1} \vec{b}}.
\end{equation}
We can easily identify this as a Clifford group operation, up to
the global phase $e^{id}$:  $\ket{\vec{z}} \mapsto
\ket{A^{-1}\vec{z}}$ can be performed with CNOT gates,
$\ket{\vec{z}} \mapsto \ket{\vec{z} - A^{-1}\vec{b}}$ can be
performed with $X$ operations, and $\ket{\vec{z}} \mapsto
e^{i\pi(\vec{c} \cdot \vec{z})} \ket{\vec{z}}$ can be performed
with $Z$ operations. \QED
\medskip

The proof for $d$-dimensional registers is almost identical,
except that we must replace $\pi$ everywhere with $2\pi/d$, and we
need scalar multiplication gates as well as SUM gates to perform
$\ket{\vec{z}} \mapsto \ket{A^{-1}\vec{z}}$ \cite{Gottesman99b}.

For prime Hilbert space dimensions there is a unique Pauli MUB construction that uses \textit{all} stabilizer states \cite{BandyopadhyayBRV02}. In that case our set of pure classical states coincides with the set of stabilizer states, and our group of classical unitaries $\{U_c\}$ coincides with the Clifford group. This is not the case for power-of-prime dimensions, where a complete set of Pauli MUB contains only a proper subset of the stabilizer states, resulting in classical unitaries $\{U_c\}$ which form a proper subgroup of the Clifford group.

The Gottesman-Knill theorem \cite{Gottesman97} states that
Clifford group operations on stabilizer states can be simulated
efficiently on a classical computer. Thus, our Theorems
\ref{thm:statics} and \ref{thm:MUBequiv} guarantee that the group
of classical unitaries $\{U_c\}$ applied on the set of classical
pure states in $C_d$ can be efficiently simulated, i.e. with a
number of time-steps that increases only polynomially in the
number of qubits. This is to be contrasted with general quantum
computation, which uses states and operations outside of our
classical sets, and which is thought to provide exponential
speedup for some problems. The necessity of negativity of $W$ for
achieving universal quantum computation had been noted in
\cite{Galvao05} for a particular computational model proposed
recently by Bravyi and Kitaev \cite{BravyiK05}.

Our results point to an interesting convergence between two
different notions of classicality. The first defines classical
states as those whose description can be made in terms of
non-negative quasi-probability distributions, in this case the
discrete Wigner functions of \cite{GibbonsHW04}. The second is
motivated by quantum computation: classical states and operations
are those which can be efficiently simulated on a classical
computer. For a related discussion of simulability in the context
of continuous variables see \cite{BartlettSBN02}.

In this section and the previous one we only made claims of classicality for \textit{pure} states in $C_d$ and their associated unitary dynamics. The
problem seems to become much more involved when we consider mixed states in
$C_d$ and their associated dynamics, which in this case will be (in general
non-unitary) completely positive maps. It is not clear whether an efficient
simulation scheme for this more general definition of classical dynamics
can be devised.

\subsection{Unitaries in $\{U_c\}$ and their action in phase space}

We have just shown that when we build a complete set of MUB using Pauli
operators, the `classical' unitaries in $\{U_c\}$ turn out to form a
subgroup of the Clifford group. The exact characterization of this subgroup
will depend on which Pauli MUB construction we pick. In this section we
restrict ourselves to the construction for $N$ qubits sketched in section \ref{sec:wignerdef},
and present some `classical' unitaries in $\{U_c\}$ together with their associated
action in phase space.
\begin{figure}
\begin{center}
\includegraphics[width=0.45\textwidth]{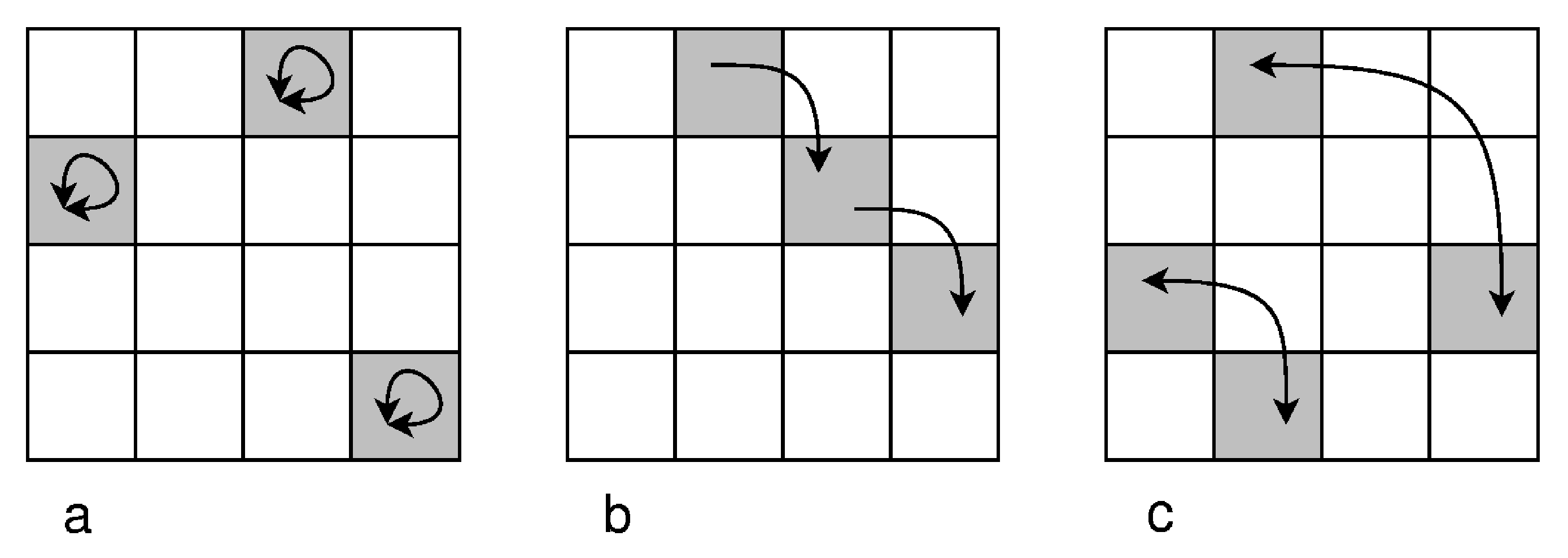}
\caption{\label{figwigner}Operators in $\{U_c\}$ map translations
$T(\vec q, \vec p)$ into $T(\vec {q'}, \vec {p'})$ (up to a sign);
representation of the transformation of translations under the
action of: a) translation operators; b) discrete squeezing; c)
Fourier transform.}
\end{center}
\end{figure}

\subsubsection{Discrete phase-space translation operators}

In \cite{PazRS04b} it was shown that the Pauli operators act as discrete
phase-space translations mapping phase-space lines into other lines. This
means that the Pauli operators themselves are in our group $\{U_c\}$.
Translation operators $\hat{T}(\vec q,\vec p)$ operate on quantum states
in such a way that their Wigner functions are transformed as flows: each
phase-space point operator is mapped into another one because of the
covariance condition imposed on the quantum net. Thus, the effect of a
translation operator on a state's Wigner function is a translation in phase
space, and in this sense its action is ``classical-like''. Since translation
operators act as flows in phase space, they preserve positivity of $W$ for any
single association between lines and MUB projectors.

\subsubsection{Discrete squeezing operator}

The discrete squeezing operator \cite{PazRS04b} maps the
horizontal and the vertical striations into themselves, while
cycling through all oblique striations. An explicit Clifford
circuit for $U_s$ is given in \cite{PazRS04b}. When acting on
translation operators, $U_s$ maps them into other translations in
a way that resembles a squeezing flow in phase space (see Fig.
1.b):
\begin{equation}
U_s \hat{T}(\vec {q}, \vec {p}) U_s^\dagger = \pm \hat{T}(\vec {q}M, \vec {p} \widetilde {M} ^{-1}).
\end{equation}

Besides covariance with respect to the discrete phase-space translations,
we can impose on the quantum net also the constraint of covariance under
$U_s$.  In doing so, the freedom in picking the quantum net will be limited to the choice of which MUB projector to associate to a fixed oblique line, since the covariance requirement determines all other associations. In this way, the number of possible choices
is greatly reduced from $d^{d-1}$ to $d$.

If we choose a quantum net which is covariant under the squeezing
operator, it can be shown that $U_s$ will map phase-space point
operators into other point operators, i.e. $U_s$ acts like a
phase-space flow. This may not be the case if the quantum net is
not chosen to be covariant under $U_s$.

By definition, the group $\{U_c\}$ consists of unitaries that preserve non-negativity of the Wigner function $W$ for all possible quantum nets. This does not imply that operators in $\{U_c\}$ will preserve positivity for any single definition of $W$. This is because some states may have positive $W$ for a single definition, but negative $W$ for other definitions (and hence lie outside the set $C_d$). In the case of $U_s$, preservation of positivity for each single definition of $W$ is only guaranteed when $U_s$ acts as a flow in phase space, and this only happens when the quantum net is chosen to be covariant with respect to $U_s$.

\subsubsection{Finite Fourier transform}

The finite Fourier transform $F$ \cite{KlimovS-SdG05} maps the
horizontal and the vertical striations into one another; oblique
striations are interchanged in pairs, and one of them [the ``main
diagonal'', which corresponds to the eigenstates of the set of
translations obtained by setting $\vec{a} = \vec {b}$ in eq.
(\ref{eqsetPaulis})] is mapped into itself. For the particular
case in which the canonical basis of the Galois field is
self-dual, $F$ is just the Hadamard transform. The effect of $F$
on the translation operators is -- up to a sign -- a reflection
with respect to the main diagonal of phase space (see Fig. 1.c):
\begin{equation}
F \hat{T}(\vec {1} M^j, \vec {1} \widetilde {M}^k) F^\dagger = \pm \hat{T}(\vec {1} M^k, \vec {1} \widetilde {M}^j)
\end{equation}

The fact that $F$ interchanges translation operators by a reflection might
suggest that its action on the states could be analogous, that is, that $F$
could reflect a state's Wigner function with respect to the main diagonal,
perhaps for some particular quantum nets (as is the case with $U_s$).

For $F$ to act on lines as a reflection with respect to the main
diagonal, there should be one MUB projector associated to the axis
of reflection, and $F$ should map this projector into itself. This
projector should, then, be a common eigenstate of $F$ and all the
Paulis that define the basis to which the state belongs. Using the
fact that $F$ anti-commutes with some of them, and that the
eigenvalues of $F$ and the Paulis are different from zero, it can
be seen that no state can fulfill this requirement. Thus, there is
no association between lines and MUB projectors that makes the
action of $F$ on the Wigner function be a reflection flow.

Moreover, it can be seen that there is no quantum net for which $F$ acts as a
flow in any way, because for $F$ to be a flow it should map phase-space point operators
into other point operators. For this to happen, the $(d+1)$ lines that
intersect in any given point must be mapped by $F$ into other lines that
intersect in only one point. The vertical and the horizontal rays (i.e. lines containing the origin) are
interchanged by $F$, so the other rays must be mapped into rays too (so
that all the resulting lines intersect at the origin). This requires the
ray in the main diagonal to be  mapped into itself, and, as pointed out in
the previous paragraph, this cannot be achieved.

Therefore, $F$ provides an example of an operator in $\{U_c\}$ which cannot
be interpreted in terms of a flow for any choice of associations between lines
and states, and so has no obvious continuous phase-space analogue.
\medskip

\section{Conclusion} \label{secdiscussion}

We have characterized the set $C_d$ of states whose discrete
Wigner functions $W$ (as defined in \cite{GibbonsHW04}) are
non-negative. We showed that the only pure states in $C_d$ are the
mutually unbiased bases projectors used to define $W$, as
conjectured in \cite{Galvao05}. Since these projectors can always
be chosen to be stabilizer states, they admit an efficient
classical description using the stabilizer formalism. Moreover, we
proved that the unitaries which preserve non-negativity of $W$ for
all such functions $W$ form a subgroup of the Clifford group. It
is known that Clifford operations on stabilizer states can be
simulated efficiently on a classical computer. We have thus
identified a relation between two different notions of
classicality: states which are classical in the sense of having
non-negative quasi-probability distributions (the discrete Wigner
functions of \cite{GibbonsHW04}) can also be simulated efficiently
on classical computers. Since general quantum computation is
thought to be hard to simulate classically, our results mean that
negativity of $W$ is \textit{necessary} for exponential
computational speedup with pure states.

There are many open problems worth investigating. The complete
characterization of non-negativity preserving unitaries for
different constructions of complete sets of MUB is still unsolved.
It would also be interesting if one could relate non-classicality
to negativity of $W$ in a \textit{quantitative} way. Another
research direction is to investigate the relationship between $W$
and a notable open problem, that of the existence of complete sets
of mutually unbiased bases for general Hilbert space dimensions
(see \cite{Bengtsson04, BengtssonE04, Wootters04b}). The original
idea behind continuous-variable Wigner functions was to help
visualize quantum dynamics in the familiar framework of classical
phase space. Some research has been done on the visualization of
quantum information protocols in discrete phase-space
\cite{MiquelPS02,PittengerR05,Wootters04,PazRS04,KoniorczykBJ01,
Paz02, LopezP03, PazRS04b}; further work might bring insights into
existing applications, or suggest new ones.

\textbf{Acknowledgments.} EFG was partly supported by Canada's
NSERC. JPP was partially funded by Fundaci\'on Antorchas and a
grant from ARDA. AOP was partially supported by NSF grants
EIA-0113137 and DMS-0309042.

\end{document}